\documentclass[12pt,a4paper]{article}
\pdfoutput=1

\usepackage{amssymb}
\usepackage{setspace}
\usepackage{indentfirst}

\begin{document}

\begin{titlepage}
\begin{center}

\vspace*{25mm}

\begin{spacing}{1.7}
{\LARGE\bf Brane SUSY Breaking and the Gravitino Mass}
\end{spacing}

\vspace*{25mm}

{\large
Noriaki Kitazawa
}
\vspace{10mm}

Department of Physics, Tokyo Metropolitan University,\\
Hachioji, Tokyo 192-0397, Japan\\
e-mail: noriaki.kitazawa@tmu.ac.jp

\vspace*{25mm}

\begin{abstract}
Supergravity models with spontaneously broken supersymmetry
 have been widely investigated over the years,
 together with some notable non--linear limits.
Although in these models the gravitino becomes naturally massive
 absorbing the degrees of freedom of a Nambu--Goldstone fermion,
 there are cases in which the naive counting of degrees of freedom does not apply, in particular
 because of the absence of explicit gravitino mass terms in unitary gauge.
The corresponding models
 require non-trivial de Sitter--like backgrounds,
 and it becomes of interest to clarify the fate of their Nambu--Goldstone modes.
We elaborate on the fact that
 these non--trivial backgrounds can accommodate, consistently, gravitino fields
 carrying a number of degrees of freedom that is intermediate
 between those of massless and massive fields in a flat spacetime.
For instance, in a simple supergravity model of this type with de Sitter background,
 the overall degrees of freedom of gravitino
 are as many as for a massive spin--$3/2$ field in flat spacetime,
 while the gravitino remains massless in the sense that
 it undergoes null--cone propagation in the stereographic picture.
On the other hand,
 in the ten--dimensional USp$(32)$ Type I Sugimoto model with ``brane SUSY breaking'',
 which requires a more complicated background,
 the degrees of freedom of gravitino are half as many of those of a massive one,
 and yet it somehow behaves again as a massless one.
\end{abstract}

\end{center}
\end{titlepage}

\section{Introduction}
\label{introduction}

If supersymmetry~\cite{susy} plays a role in the fundamental interactions,
 one is naturally led to Supergravity~\cite{sugra} and String Theory~\cite{stringtheory},
 where supersymmetry must be spontaneously broken at low energies.
There is no general agreement, as of today,
 on the detailed dynamics of supersymmetry breaking in String Theory,
 and therefore it is important to explore further
 its model--independent realizations in the low--energy effective Supergravity.
Non--linear realizations of supersymmetry are particularly interesting,
 because they encode information that is independent of the detailed fundamental dynamics.

The non-linear realization of global supersymmetry emerged very early,
 in the Volkov-Akulov model~\cite{Volkov:1973ix},
 whose first application to supergravity models was discussed
 in~\cite{Volkov:1973jd,Volkov:1974ai,Deser:1977uq}.
This framework remains of great interest,
 and its coupling to Supergravity was recently reconsidered in~\cite{vaconstraint},
 in the light of constrained superfields~\cite{constrained},
 and in the light of non-BPS D-branes~\cite{Bandos:2015xnf,Bandos:2016xyu}.
As was explained in \cite{Deser:1977uq},
 one must introduce mass terms for the gravitino and Nambu--Goldstone fermion
 and modify accordingly its supersymmetry transformation
 in order to eliminate a cosmological constant term to arrive at flat--space models.
The systematics of these constructions, examined in detail in~\cite{cfgvp},
 led eventually to the no--scale models of~\cite{noscale},
 which also found a string realization in the presence of internal fluxes~\cite{fluxes}.

Typically,
 the gravitino becomes massive absorbing the degrees of freedom of a Nambu--Goldstone fermion,
 a phenomenon that becomes manifest in a unitary gauge,
 but in the ten--dimensional and six--dimensional orientifold models~\cite{orientifolds}
 with ``brane supersymmetry breaking''~\cite{Sugimoto:1999tx,bsb}
 supersymmetry is non--linearly realized
 and \emph{no explicit gravitino mass term is allowed}~\cite{Dudas:2000nv},
 since the gravitino is a Majorana-Weyl fermion.\footnote{
``Brane supersymmetry breaking''
 is a way to break supersymmetry by brane configurations without tachyon instabilities.
For example,
 the simultaneous presence of branes and anti-branes in order to break supersymmetry
 gives rise to tachyon instabilities corresponding to their pair annihilations.
The combination of anti-D-branes and USp-type orientifold fixed planes
 is a typical configuration of ``brane supersymmetry breaking''
 in which no tachyon instability appears,
 because these two objects do not annihilate and the system can be stable.
See the review article in \cite{bsb} for further details.}
On the other hand,
 these models require non--trivial backgrounds,
 since they include a dilaton--dependent cosmological term~\cite{Dudas:2000ff}.
Since no gravitino mass term emerges in a unitary gauge,
 the role of the Nambu--Goldstone fermion may appear confusing.
Our aim here is
 to elaborate on the fate of the degrees of freedom of Nambu--Goldstone fermions
 in this second class of models.
There are some interesting aspects in this story,
 since the curved spacetime is crucial in a proper account of the related degrees of freedom.

Much effort was devoted, over the years,
 to providing suitable definitions of masses and degrees of freedom in de Sitter spacetime,
 in particular in~\cite{Deser:1983mm,mass-in-deSitter,Deser:2004ji}.
A highlight of these works is that null--cone propagation
 takes naturally the place of masslessness in flat space time~\cite{Deser:1983mm},
 and the correspondence is illuminated by the special choice of
 ``symmetric coordinates'' described in~\cite{Gursey:1963ir}.
The detailed investigation of this criterion of masslessness in the second class of models
 will lead us to a variant that is more suitable for ``brane supersymmetry breaking'',
 a ``cosmological'' criterion of masslessness in flat slicing coordinates.
The new criterion applies, in particular,
 to the ten--dimensional Sugimoto model of \cite{Sugimoto:1999tx}
 whose background is not exactly de Sitter spacetime.
As we shall see,
 the gravitino in the Sugimoto model remains surprising massless in this sense,
 although it does absorb the degrees of freedom of a Nambu--Goldstone fermion.

In the next section
 we recall briefly some aspects of non--linear supersymmetry and the super--Higgs mechanism.
As a warm up,
 we compare two simple models of pure supergravity without matter fields
 with and without a gravitino mass term.
The latter model is
 an explicit realization of the equations and constraints for a massless spin--$3/2$ field
 that were described in~\cite{Deser:1983mm} without specifying the action.
This is an instructive step,
 since the low--energy supergravity action of the Sugimoto model
 is also an example of non--linear supersymmetry without a gravitino mass term.
In section~\ref{massless-in-dS} we discuss
 the massless condition for fermions in a de Sitter spacetime.
We briefly review
 the role of null--cone propagation in the symmetric coordinate system and conformal invariance,
 investigating explicitly a simple model involving a spinor field.
We then propose a new masslessness criterion for fermion fields,
 which refers to the ``rest mass'' contribution to the ``cosmological'' energy density
 in a flat slicing coordinate system for a de Sitter spacetime.
In section \ref{massless-gravitino}
 we demonstrate that,
 in a simple model without gravitino mass term that will be introduced in the next section,
 the gravitino remains massless in this sense,
 displaying its null--cone propagation and also applying our new criterion.
We then show that
 the gravitino in the Sugimoto model remains massless,
 in the sense specified above, via our new criterion,
 which applies insofar as the background can be understood as a cosmological evolution.
The overall number of degrees of freedom of the massless gravitino
 is the sum of those of a massless gravitino in flat spacetime and a Nambu--Goldstone fermion,
 which is half as many of those of a massive spin--$3/2$ field.
The last section contains some concluding remarks.

\section{Supergravity models with non-linear supersymmetry}
\label{SUGRA-NL-SUSY}

Let us begin by considering pure supergravity
\begin{equation}
 {\cal L}_{\rm pure}
  = \frac{1}{2\kappa^2} \, e \, {\cal R}(e,\omega)
  - e \, \frac{1}{2} {\bar \psi}_\mu \Gamma^{\mu\rho\nu} D_\rho \psi_\nu + {\cal O}((\psi_\mu)^4),
\end{equation}
 while leaving aside higher powers in the Majorana gravitino field.
Here, we work sketchily in a generic spacetime dimension $D \geq 4$,
 which allows Majorana fermions,
 at the cost of leaving aside some members of the supergravity multiplet.
In this paper we follow the conventions of~\cite{Tanii:2014gaa}.
This Lagrangian is invariant,
 up to a total divergence and higher powers of gravitino, under the transformations
\begin{eqnarray}
 \delta e^m{}_\mu &=& - \frac{1}{2} \kappa \, {\bar \psi}_\mu \Gamma^m \epsilon,
\\
 \delta \psi_\mu &=& \frac{1}{\kappa} D_\mu \epsilon,
\end{eqnarray}
 whose parameter is the Majorana spinor $\epsilon$.
The covariant derivative of the spinor field involves the spin connection $\omega$ and reads
\begin{equation}
 D_\mu \epsilon =
  \left(
   \partial_\mu + \frac{1}{4} \omega_\mu{}^{mn} \Gamma_{mn}
  \right) \epsilon,
\end{equation}
 where local Lorentz indices are denoted by Latin letters.
Let us now introduce a Nambu--Goldstone fermion field $\theta$
 that provides a non--linear realization of supersymmetry
\begin{equation}
 {\cal L}_{\rm NL}
 = - e \, 2 f^2 - e \, \frac{1}{2} {\bar \theta} \Gamma^\mu D_\mu \theta
   + {\cal O}((\theta)^4),
\end{equation}
 whose lowest--order coupling to the gravitino
\begin{equation}
 {\cal L}_{\rm current} = e \, \kappa {\bar \psi}_\mu S^\mu
\end{equation}
 involves the supersymmetry current
\begin{equation}
 S_\mu = - f \Gamma_\mu \theta \ .
\end{equation}
The dimensionful quantity $f$ defines the scale of supersymmetry breaking
 and the Nambu--Goldstone fermion field transforms, to lowest order, according to
\begin{equation}
 \delta \theta = f \epsilon \ .
\end{equation}

The total Lagrangian ${\cal L}_{\rm pure}+{\cal L}_{\rm NL}+{\cal L}_{\rm current}$
\begin{equation}
 {\cal L}
  = \frac{1}{2\kappa^2} \, e \, {\cal R}(e,\omega)
  - e \, \frac{1}{2} {\bar \psi}_\mu \Gamma^{\mu\rho\nu} D_\rho \psi_\nu
  - e \, \frac{1}{2} {\bar \theta} \Gamma^\mu D_\mu \theta
  + e \, \kappa {\bar \psi}_\mu S^\mu
  - e \, 2 f^2
\label{simple-model}
\end{equation}
 is invariant under these supersymmetry transformations,
 up to higher terms that we are not tracking here~\cite{freedman,Deser:1977uq}.
This example involves on purpose no gravitino mass term in the unitary gauge $\theta=0$.
Note that this model contains a positive cosmological constant term,
 so that one is naturally led to consider a background de Sitter spacetime.
The lesson of these old considerations is that the absence of the mass term
 does not necessarily imply that the gravitino is a simple massless spin--$3/2$ field,
 which conforms to the fate of the degrees of freedom introduced by Nambu--Goldstone fermion.
The purpose of this paper is
 to elaborate on what happens to the gravitino and the Nambu--Goldstone fermion
 in more complicated models of broken Supersymmetry.

Let us now turn to another pure supergravity model
 with a gravitino mass and a negative cosmological constant, so that
\begin{equation}
 {\cal L}_{\rm pure (massive)}
  = \frac{1}{2\kappa^2} \, e \, {\cal R}(e,\omega)
  - e \, \frac{1}{2} {\bar \psi}_\mu \Gamma^{\mu\rho\nu} D_\rho \psi_\nu
  + e \, \frac{1}{2} m {\bar \psi}_\mu \Gamma^{\mu\nu} \psi_\nu
  + e \, C,
\end{equation}
 where
\begin{equation}
 C = \frac{2(D-1)}{(D-2)} \frac{m^2}{\kappa^2}.
\end{equation}
The gauge transformations of this model are different, and include the contributions
\begin{eqnarray}
 \delta e^m{}_\mu &=&
  - \frac{1}{2} \kappa \, {\bar \psi}_\mu \Gamma^m \epsilon,
\\
 \delta \psi_\mu &=&
  \frac{1}{\kappa} \left( D_\mu \epsilon + \frac{1}{D-2} m \Gamma_\mu \epsilon \right).
\end{eqnarray}
Making the Lagrangian ${\cal L}_{\rm NL}+{\cal L}_{\rm current}$
 invariant under the new gauge transformation
 requires the introduction of a mass term for the Nambu--Goldstone fermion~\cite{zumino},
\begin{equation}
 {\cal L}_{\rm mass} = - e \, \frac{1}{2} \frac{D}{D-2}\, m \,{\bar \theta} \theta.
\end{equation}
The portion of the total Lagrangian that we are focussing on,
 ${\cal L}_{\rm pure(masive)}+{\cal L}_{\rm NL}+{\cal L}_{\rm current}+{\cal L}_{\rm mass}$,
 is then invariant under the new gauge transformation,
 up to higher order terms or
 contributions involving the other members of the gravity or matter multiplets.
In particular,
 a model without cosmological constant term would require tuning $2f^2=C$,
 and thus linking the mass parameters of the gravitino and the Nambu--Goldstone fermion, $m$,
 to the supersymmetry breaking scale $f$.
In a model of this type admitting a flat spacetime background
 the gravitino has a mass term in the unitary gauge $\theta=0$,
 and in fact it is a conventional massive spin--$3/2$ field.

Models that are similar to some extent to those in eq.(\ref{simple-model})
 are realized in orientifolds with brane supersymmetry breaking,
 and the ten--dimensional Sugimoto model is the simplest example in this class.
At tree level
 the supersymmetry present in the original type IIB closed string sector
 is halved by the orientifold projection, as in the type--I superstring,
 while in the open sector supersymmetry is completely broken, or non--linearly realized.
The low--energy effective Lagrangian
 of the Sugimoto model was discussed in detail in~\cite{Dudas:2000nv}.
In Einstein frame and \emph{in unitary gauge} it combines open and closed contributions,
\begin{equation}
 {\cal L} = {\cal L}_{\rm closed} + {\cal L}_{\rm open},
\end{equation}
 where
\begin{eqnarray}
 {\cal L}_{\rm closed}
 &=&
  \frac{1}{2\kappa_{10}^2}
  e \,
  \Bigg\{
   R
   - \frac{1}{2} \partial^\mu \phi \partial_\mu \phi
   - \frac{1}{12} e^{-\phi} F^{\mu\nu\rho} F_{\mu\nu\rho}
   - \frac{1}{2} {\bar \psi}_\mu \Gamma^{\mu\rho\nu} D_\rho \psi_\nu
   - \frac{1}{2} {\bar \lambda} \Gamma^\mu D_\mu \lambda
\nonumber\\
   &-& \frac{1}{2\sqrt{2}} {\bar \psi}_\mu \Gamma^\nu \Gamma^\mu \lambda \partial_\nu \phi
   - \frac{1}{48} e^{- \frac{1}{2} \phi}
   \left[
    {\bar \psi}_{[\mu} \Gamma^\mu \Gamma^{\rho\sigma\tau} \Gamma^\nu \psi_{\nu]}
    - \sqrt{2} {\bar \psi}_\mu \Gamma^{\rho\sigma\tau} \Gamma^\mu \lambda
   \right] F_{\rho\sigma\tau}
  \Bigg\}
\end{eqnarray}
 describes the contribution from the closed sector, while
\begin{eqnarray}
 {\cal L}_{\rm open}
 &=&
  \frac{1}{2\kappa_{10}^2}
  e \,
  \Bigg\{
   - e^{\frac{1}{2} \phi} \frac{1}{2g^2} {\rm tr} \left( F^{\mu\nu} F_{\mu\nu} \right)
   - \frac{1}{g^2} {\rm tr} \left( {\bar \chi} \Gamma^\mu D_\mu \chi \right)
\nonumber\\
 && \qquad\qquad
   - 2 \alpha_E e^{\frac{3}{2} \phi}
   + e^{- \frac{1}{2} \phi} \frac{1}{24 g^2}
     {\rm tr} \left( {\bar \chi} \Gamma^{\rho\sigma\tau} \chi \right) F_{\rho\sigma\tau}
  \Bigg\}
\end{eqnarray}
 describes the contribution from open sector.
Beginning from the closed sector,
 $\phi$ is the dilaton field,
 $F_{\mu\nu\rho}$ is the field strength tensor of an antisymmetric tensor field
 (a Ramond--Ramond field),
 $\psi_\mu$ is the gravitino field and $\lambda$ is the dilatino field.
Turning now to the open sector,
 $F_{\mu\nu}$ is the field strength of the USp$(32)$ gauge fields and
 $\chi$ is a spinor field
 belonging to the \emph{traceless} anti--symmetric representation of the USp$(32)$ gauge group.
On the other hand the Nambu--Goldstone fermion,
 which has been set to zero in the unitary gauge,
 would arise from the corresponding symplectic trace.
The gauge coupling and ``cosmological constant'' are
 $g^2 = 2/\alpha'$ and $\alpha_E = 64 T_9 \kappa_{10}^2$, respectively,
 where $T_9$ is the tension of anti-D$9$-brane.

In the following we also set to zero
 the fields $F_{\mu\nu\rho}$, $F_{\mu\nu}$ and $\chi$,
 since they do not play any role in the non--trivial backgrounds of interest
 that are required by the dilaton--dependent cosmological term
 and in the corresponding realization of supersymmetry breaking.
The model thus reduces to
\begin{equation}
 {\cal L}
 = \frac{1}{2\kappa_{10}^2}
   e \,
   \Bigg\{
    R
    - \frac{1}{2} \partial^\mu \phi \partial_\mu \phi
    - \frac{1}{2} {\bar \psi}_\mu \Gamma^{\mu\rho\nu} D_\rho \psi_\nu
    - \frac{1}{2} {\bar \lambda} \Gamma^\mu D_\mu \lambda
    - \frac{1}{2\sqrt{2}} {\bar \psi}_\mu \Gamma^\nu \Gamma^\mu \lambda \partial_\nu \phi
    - 2 \alpha_E e^{\frac{3}{2} \phi}
   \Bigg\}.
\label{model-Sugimoto}
\end{equation}
Now the field equation of the Nambu--Goldstone fermion yields
\begin{equation}
 \Gamma^\mu \psi_\mu = - \frac{3}{\sqrt{2}} \lambda,
\label{constraint-Sugimoto}
\end{equation}
 which identifies the dilatino and the gamma trace of the gravitino.
We see that there is no mass term for gravitino,
 and therefore this model belongs to the same class as the model of eq.(\ref{simple-model}).
There is an additional reason for the absence of a gravitino mass term:
 in this ten--dimensional setting this field is a Majorana--Weyl fermion field,
 for which the mass term simply does not exist.
The goal of this paper is
 to elaborate on what happens to the gravitino in this more complicated model,
 in the presence of non--trivial background fields.

\section{Massless propagation in de Sitter backgrounds}
\label{massless-in-dS}

It is well known that
 one can define a $D$--dimensional de Sitter spacetime starting from the quadratic constraint
\begin{equation}
 \eta_{\alpha\beta} \xi^\alpha \xi^\beta
  = - (\xi^0)^2 + (\xi^1)^2 + \cdots + (\xi^D)^2 = 1/H^2 \ .
\label{dS-ambient}
\end{equation}
Here $\xi^\alpha$ with $\alpha = 0,1, \cdots, D$ are ambient coordinates
 and $1/H$ is the de Sitter radius,
 which is related to the cosmological constant $\Lambda > 0$ according to
 $\Lambda = (D-1)(D-2) H^2 / 2\kappa^2$.
The curvature tensors are described
 by the metric tensor $g_{\mu\nu}$, independently of the coordinate system, as
\begin{equation}
 R^\rho{}_{\mu\sigma\nu}
  = H^2 \left( \delta^\rho_\sigma g_{\mu\nu} - \delta^\rho_\nu g_{\mu\sigma} \right)\ ,
\qquad
 R_{\mu\nu} = R^\rho{}_{\mu\rho\nu} = (D-1) H^2 g_{\mu\nu} \ ,
\end{equation}
 with $\mu, \nu, \rho, \sigma = 0,1,\cdots, D-1$,
 and the scalar curvature is a constant, with $R = D (D-1) H^2$.
In this paper use two concrete coordinate systems,
 the symmetric coordinates and the flat slicing coordinates.

The symmetric coordinates \cite{Gursey:1963ir},
 the result from a stereographic projection of
 the generalized hyperboloid of eq.(\ref{dS-ambient}) to $D$-dimensional flat spacetime,
 defined via
\begin{equation}
 \xi^\mu = \frac{x^\mu}{1+s},
\qquad
 \xi^D = \frac{1-s}{1+s} \frac{1}{H},
\end{equation}
 where
\begin{equation}
 s = \frac{H^2}{4} \eta_{\mu\nu} x^\mu x^\nu.
\end{equation}
In this fashion the metric tensor takes the form
\begin{equation}
 g_{\mu\nu} = \Omega^2 \eta_{\mu\nu},
\label{metric-symmetric}
\end{equation}
 with $\Omega = (1+s)^{-1}$, while
\begin{equation}
 e_\mu{}^m = \Omega \delta_\mu^m,
\qquad
 e_m{}^\mu = \Omega^{-1} \delta_m^\mu
\end{equation}
 is a corresponding vielbein.
The Christoffel symbols and the spin connection are then
\begin{equation}
 \Gamma^\rho{}_{\mu\nu}
 = \frac{\Omega H^2}{2}
   \left[
    x^\rho \eta_{\mu\nu} - x_\mu \delta^\rho_\nu - x_\nu \delta^\rho_\mu
   \right],
\end{equation}
\begin{equation}
 \frac{1}{4} \omega_\mu{}^{mn} \Gamma_{mn} = - \Omega \frac{H^2}{4} \gamma_{\mu\nu} x^\nu,
\end{equation}
 respectively, where
\begin{equation}
 \Gamma^\mu = \gamma^m e_m{}^\mu = \Omega^{-1} \gamma^\mu,
\qquad
 \{\gamma^m, \gamma^n\} = 2 \eta^{mn}.
\end{equation}
Since this coordinate system yields a conformally flat metric,
 it is a convenient choice to investigate null--cone propagation and conformal covariance
 of the field equations.

The more conventional flat slicing coordinates of de Sitter spacetime are defined as
\begin{equation}
 \xi^0 = \frac{1}{H} \sinh(Hx^0) + \frac{H}{2} r^2 e^{Hx^0},
\qquad
 \xi^i = x^i e^{Hx^0},
\qquad
 \xi^D = \frac{1}{H} \cosh(Hx^0) - \frac{H}{2} r^2 e^{Hx^0},
\end{equation}
 where $r^2=\delta^{ij} x^i x^j$ with $i,j = 1, \cdots, D-1$.
The metric is then
\begin{equation}
 ds^2
 = g_{\mu\nu} dx^\mu dx^\nu
 = - (dx^0)^2 + e^{2 H x^0} \delta^{ij} dx^i dx^j,
\end{equation}
 and a corresponding vielbein is
\begin{equation}
 e_\mu{}^m = {\rm diag} \left( 1, e^{H x^0}, e^{H x^0}, \cdots \right).
\end{equation}
Christoffel symbols and spin connection are now
\begin{equation}
 \Gamma^0{}_{ij} = H g_{ij},
\qquad
 \Gamma^i{}_{0j} = \Gamma^i{}_{j0} = H g^i_j,
\qquad
 \mbox{others = 0},
\end{equation}
 and
\begin{equation}
 \omega_\mu{}^{0{\bar m}} = - \omega_\mu{}^{{\bar m}0} = H e_\mu{}^{{\bar m}},
\qquad
 \mbox{others = 0},
\end{equation}
 respectively, where ${\bar m} = 1,2, \cdots, D-1$.
As is well known,
 this coordinate system covers half of the whole de Sitter spacetime,
 the portion where space expands in time, which affords a natural cosmological interpretation.

In a Minkowski spacetime
 the Casimir operator $\eta_{\mu\nu} \,\hat{p}^\mu \hat{p}^\nu$ of Poincar\'e group,
 where $\hat{p}^\mu$ is the generator of spacetime translation,
 is used to define the mass of the field.
Since there is no such Casimir operator in the de Sitter SO$(D,1)$ group,
 there is no natural way to define the mass of fields.
Still, null--cone propagation remains a convincing criterion for masslessness.

Let us now briefly recall
 the massless criterion via null--cone propagation
 in the symmetric coordinate system \cite{Deser:1983mm}.
To this end,
 let us consider a simple spin--$1/2$ spinor field in de Sitter background, for which
\begin{equation}
 {\cal L}_{\rm spin \, 1/2} = - e \, {\bar \Psi} \Gamma^\mu D_\mu \Psi.
\label{model-spin-1/2}
\end{equation}
The field equation
\begin{equation}
 \Gamma^\mu D_\mu \Psi = 0
\end{equation}
 can be rewritten as
\begin{equation}
 \Gamma^\mu D_\mu \Psi = \Omega^{-\frac{D+1}{2}} \gamma^\mu \partial_\mu \psi = 0.
\end{equation}
 where $\psi$ is the rescaled field defined by
\begin{equation}
 \Psi = \Omega^{w_{1/2}} \psi
\label{rescale-1/2}
\end{equation}
 with the conformal weight of the spinor field $w_{1/2} = (1-D)/2$.
One is thus led to null--cone propagation,
 and in this respect the spinor field can be considered a massless field in de Sitter background.
A similar conclusion can be reached starting from the second--order field equation.
\begin{equation}
 \Gamma^\mu D_\mu \Gamma^\nu D_\nu \Psi
 = \left(
    g^{\mu\nu} D_\mu D_\nu - \frac{D(D-1)}{4} H^2
   \right) \Psi
 = \Omega^{-\frac{3+D}{2}} \eta^{\mu\nu} \partial_\mu \partial_\nu \Psi = 0
\end{equation}
 which leads consistently to null--cone propagation
 and thus points again to the notion of a  massless spinor field.

The action of this model is invariant under the de Sitter SO$(D,2)$ conformal transformation.
Since local Weyl invariance is enough for de Sitter conformal invariance \cite{Deser:2004ji},
 we show local Weyl invariance of this model.
The Weyl transformation is defined as
\begin{equation}
 e_\mu{}^m \longrightarrow \Omega_W e_\mu{}^m,
\qquad
 e_m{}^\mu \longrightarrow \Omega_W^{-1} e_\mu{}^m,
\end{equation}
 to be combined with
\begin{equation}
 \Psi \longrightarrow \Omega_W^{w_{1/2}} \Psi,
\end{equation}
 where $\Omega_W$ is the local Weyl scaling factor.
Using the definition of the gamma matrices in a non--trivial background,
 $\Gamma^\mu = \gamma^m e_m{}^\mu$, and the transformation of spin connection
\begin{equation}
 \omega_\mu{}^{mn} \longrightarrow
  \omega_\mu{}^{mn}
   + \left( e^m{}_\mu e^{n\nu} - e^n{}_\mu e^{m\nu} \right) \partial_\nu \ln \Omega_W,
\end{equation}
 it is straightforward to see the invariance of the Lagrangian of eq.(\ref{model-spin-1/2})
 provided one chooses the conformal weight $w_{1/2} = (1-D)/2$.
Since conformal invariance reflects the absence of a specific scale,
 it is natural to regard the spinor field as a massless field.

We now turn to propose
 a massless criterion for fermion fields in the flat slicing coordinate system,
 referring again to the above model of a spinor field.
The explicit form of the field equation in the flat slicing coordinate is
\begin{equation}
 \partial_0 \Psi \ + \ \frac{D-1}{2} H \Psi = 0.
\end{equation}
Here, we have also assumed that the field depends only on time,
 since we anticipate the use of some physical arguments
 related to the homogeneous expansion of the Universe.
The important property of the solution of this equation is that it is not oscillatory,
\begin{equation}
 \Psi \propto e^{-\frac{D-1}{2} H x^0},
\label{solution-1/2}
\end{equation}
 and consequently does not contribute to the energy density $T_{00}$, where
\begin{equation}
 T_{\mu\nu}
  = \frac{1}{2}
    \left(
     {\bar \Psi} \Gamma_\nu D_\mu \Psi - {\bar \Psi} \overleftarrow{D}_\mu \Gamma_\nu \Psi
    \right).
\end{equation}
The explicit form of $T_{00}$ in our present setting
\begin{equation}
 T_{00}
  = \frac{1}{2}
    \left\{
     \Psi^\dag ( i \partial_0 \Psi ) - ( i \partial_0 \Psi^\dag) \Psi
    \right\}
\end{equation}
 clearly indicates that it could be non--vanishing for an oscillatory solution
 $\Psi \propto e^{-i m x^0}$ with ``rest mass'' $m$.
Hence,
 the non--oscillatory behavior of the Fermi field can regarded as an indication of masslessness.

We can now propose a similar argument for the second--order field equation,
 anticipating some steps that will prove useful for the application to the gravitino
 in the next section.
The explicit form of the second--order field equation is
\begin{equation}
 \partial_0 \partial_0 \Psi + (D-1) H \partial_0 \Psi
  + \left( \frac{D-1}{2} \right)^2 H^2 \Psi = 0.
\label{second-oeder-eq-1/2}
\end{equation}
Let us now perform the rescaling of the field as in eq.(\ref{rescale-1/2})
 with $\Omega = e^{H x^0}$.
The reason for this choice of $\Omega$ is that
 the metric can be described, in conformal time $\eta$, as
\begin{equation}
 ds^2 = \Omega^2 \left( -d\eta^2 + \delta^{ij} dx^i dx^j \right)
\end{equation}
 with $\Omega = 1/\vert H \eta \vert = e^{H x^0}$,
 so that this setting compares naturally with eq.(\ref{metric-symmetric}).
The field equation now takes the simple form
\begin{equation}
 \partial_0 \partial_0 \psi = 0.
\end{equation}
If we assume that a solution should be finite in the limit of $x^0 \rightarrow \infty$,
 $\psi$ should be a constant, which is consistent with eq.(\ref{solution-1/2}).
In this respect,
 a constant solution for the rescaled field in the flat slicing coordinate system
 can be a criterion of masslessness.

\section{Massless Gravitino in non--linear supersymmetry}
\label{massless-gravitino}

Consider the model of eq.(\ref{simple-model})
 in the background de Sitter spacetime required by the cosmological constant $2f^2$.
The field equations of the gravitino and the Nambu--Goldstone fermion are
\begin{equation}
 \Gamma^{\mu\rho\nu} D_\rho \psi_\nu + \kappa f \Gamma^\mu \theta = 0,
\end{equation}
\begin{equation}
 \Gamma^\mu D_\mu \theta - \kappa f \Gamma^\mu \psi_\mu = 0.
\end{equation}
In the unitary gauge $\theta=0$,
 these equations reduce to one field equation with one constraint:
\begin{equation}
 \Gamma^{\mu\rho\nu} D_\rho \psi_\nu = 0,
\qquad
 \Gamma^\mu \psi_\mu = 0,
\end{equation}
 the combination that was discussed as a model of null--cone propagation
 without de Sitter conformal invariance in \cite{Deser:1983mm}.
Applying $\Gamma_\mu$ to the field equation gives another constraint
\begin{equation}
 D_\mu \psi^\mu = 0
\end{equation}
 and if $D \ne 2$ the field equation becomes
\begin{equation}
 \Gamma^\mu D_\mu \psi_\nu = 0.
\end{equation}

In the symmetric coordinate system
 the explicit form of this field equation for the rescaled field
\begin{equation}
 \psi_\mu \longrightarrow \Omega^{w_{3/2}} \psi_\mu \ ,
\end{equation}
 with $w_{3/2}=(3-D)/2$,  is now
\begin{equation}
 \Omega^{-\frac{D-1}{2}} \gamma^\mu \partial_\mu \psi_\nu
  - \Omega^{-\frac{D-3}{2}} \frac{H^2}{2}
    \left(
     \gamma_\nu x^\mu \psi_\mu - x_\nu \gamma^\mu \psi_\mu
    \right) = 0.
\end{equation}
Eliminating $\gamma^\mu \psi_\mu$ and $x^\mu \psi_\mu$
 using the explicit forms of the two constraints
\begin{equation}
 \gamma^\mu \psi_\mu = 0,
\qquad
 \partial_\mu \psi^\mu = \Omega \frac{DH^2}{4} x^\mu \psi_\mu,
\end{equation}
 yields finally the field equation
\begin{equation}
 \gamma^\mu \partial_\mu \psi_\nu - \frac{2}{D} \gamma_\nu \partial_\mu \psi^\mu = 0.
\label{conformally-covariant-field-eq}
\end{equation}
This is a well--known conformally covariant field equation.
It was described in \cite{Deser:1983mm},
 and was derived in \cite{Barut:1981mt} in $D=4$.
The difference with respect to the massless Rarita--Schwinger equation
 lies in the special value of the coefficient of the second term,
 which would be one for the massless Rarita--Schwinger equation rather than $2/D$.
Note that one can not take a naive flat limit $H \rightarrow 0$
 in this conformally covariant field equation,
 since in the process to obtain it we had to perform a division by $H^2$
 in order to eliminate the $x^\mu \psi_\mu$ term.
As a result, the number of degrees of freedom of this spin--$3/2$ field
 undergoes a discontinuity in moving between flat and de Sitter spacetimes.

Acting on eq.(\ref{conformally-covariant-field-eq})
 with the ``flat'' divergence $\partial^\nu$ now gives, for $D \neq 2$,
\begin{equation}
 \gamma^\mu \partial_\mu (\partial_\nu \psi^\nu) = 0,
\end{equation}
 so that the component $\partial_\nu \psi^\nu$ carrying spin--$1/2$ degrees of freedom
 propagates on the null cone.
As a result, the other components of the field ${\tilde \psi}_\mu$,
 with $\partial_\mu {\tilde \psi}^\mu = 0$ (and also $\gamma^\mu {\tilde \psi}_\mu = 0$)
 satisfy the field equation
\begin{equation}
 \gamma^\mu \partial_\mu {\tilde \psi}_\nu = 0
\end{equation}
 which means again null--cone propagation.
Therefore,
 the gravitino of this model can be regarded as massless in de Sitter spacetime,
 although its degrees of freedom
 combine those of a massless spin--$3/2$ field and of a massless spin--$1/2$ field
 in a flat Minkowski background.
Let us stress that
 this is a different state of affairs from the ``partially massless field''
 in \cite{Deser:2001us}.

In the flat slicing coordinate system,
 under the assumption that the field depends only on time,
 the second order field equation before rescaling the field
\begin{equation}
 \Gamma^\mu D_\mu \Gamma^\nu D_\nu \psi_\rho
 = g^{\mu\nu} D_\mu D_\nu \psi_\rho
 - \frac{1}{2} R^\lambda{}_{\rho\mu\nu} \Gamma^{\mu\nu} \psi_\lambda
 - \frac{1}{4} R \psi_\rho = 0
\end{equation}
 takes the form
\begin{equation}
 \partial_0 \partial_0 \psi_0 + (D-1) H \partial_0 \psi_0
  + \frac{(D-5)(D-1)}{4} H^2 \psi_0 = 0,
\end{equation}
\begin{equation}
 \partial_0 \partial_0 \psi_i + (D-3) H \partial_0 \psi_i
  + \frac{(D-5)(D-1)}{4} H^2 \psi_i
  + H^2 \Gamma_{0i} \psi_0 - H^2 \Gamma_i{}^\lambda \psi_\lambda = 0 \ .
\end{equation}
The constraint $D_\mu \psi^\mu = 0$ now reads
\begin{equation}
 \partial_0 \psi_0 - \frac{1}{2} H \Gamma_0 \Gamma^i \psi_i + (D-1) H \psi_0 = 0,
\end{equation}
 and a simple solution of the above constraint and $\Gamma^\mu \psi_\mu = 0$ is
\begin{equation}
 \psi_0 = 0,
\qquad
 \Gamma^i \psi_i = 0.
\end{equation}
In this fashion the second--order field equation becomes
\begin{equation}
 \partial_0 \partial_0 \psi_i + (D-3) H \partial_0 \psi_i
  + \left( \frac{D-3}{2} \right)^2 H^2 \psi_i = 0.
\end{equation}
This equation is very similar to eq.(\ref{second-oeder-eq-1/2}) for the spinor field.
Rescaling the field according to
\begin{equation}
 \psi_i \longrightarrow \Omega^{w_{3/2}} \psi_i,
\end{equation}
 where $\Omega = e^{Hx^0}$
 and $w_{3/2} = (3-D)/2$ is the conformal weight of vector-spinor field,
 the field equation takes finally the very simple form
\begin{equation}
 \partial_0 \partial_0 \psi_i = 0.
\end{equation}
If we add the reasonable assumption that
 the solution should be finite in the limit of $x^0 \rightarrow \infty$,
 $\psi_i$ can only be a constant.
This means that gravitino is massless,
 exactly as was the case for the spinor field in previous section.
The energy-momentum tensor of the field before rescaling can be recast in the form
\begin{equation}
 T_{\mu\nu}
 = \frac{1}{4}
   g_{\rho\sigma}
   \left(
    {\bar \psi}^\rho \Gamma_\nu D_\mu \psi^\sigma
     - {\bar \psi}^\rho \overleftarrow{D}_\mu \Gamma_\nu \psi^\sigma
   \right)
 - \frac{1}{4}
   \left(
    {\bar \psi}^\rho \Gamma_\nu D_\rho \psi_\mu
     - {\bar \psi}_\mu \overleftarrow{D}_\rho \Gamma_\nu \psi^\rho
   \right)
\end{equation}
 and the explicit form of $T_{00}$ in our present setting is
\begin{equation}
 T_{00}
 = \frac{1}{4}
   g_{ij}
   \left\{
    (\psi^i)^\dag (i \partial_0 \psi^j) - (i \partial_0 (\psi^i)^\dag) \psi^j
   \right\}.
\end{equation}
We see that this energy density function vanishes for the non--oscillatory solutions,
 as pertains to fields with a vanishing ``rest mass''.

We are now ready
 to investigate the behavior of the gravitino in the more complicated Sugimoto model.
The relevant starting point was already introduced at the end of Section \ref{SUGRA-NL-SUSY},
 in eqs.(\ref{model-Sugimoto}) and (\ref{constraint-Sugimoto}).
A time-dependent background, which can be interpreted as a cosmological evolution,
 was first obtained in \cite{Dudas:2000ff} starting from the ansatz
\begin{equation}
 ds^2 = - e^{2B} (dx^0)^2 + e^{2A} \delta^{ij} dx^i dx^j,
\end{equation}
 for the metric, where $A$ and $B$ are functions only of $x^0$.
The gauge condition $B= - 3 \phi/4$
 which corresponds to a convenient choice for the time coordinate
 reduces the equations for the background to
\begin{equation}
\left\{
\begin{array}{l}
 \ddot{\phi} + (9 \dot{A} - \dot{B}) \dot{\phi} = -3, \\
 8 \ddot{A} - 8 \dot{A} \dot{B} + 36 ( \dot{A} )^2 + \frac{1}{4} ( \dot{\phi} )^2 = 1, \\
 36 ( \dot{A} )^2 - \frac{1}{4} ( \dot{\phi} )^2 = 1,
\end{array}
\right.
\end{equation}
 where dots indicate derivatives
 with respect to the dimensionless time $\tau = \sqrt{\alpha_E} x^0$.
The solution then reads
\begin{equation}
\left\{
\begin{array}{l}
 A = A_0 + \frac{1}{18} \ln \tau + \frac{1}{16} \tau^2, \\
 \phi = \phi_0 + \frac{2}{3} \ln \tau - \frac{2}{4} \tau^2. \\
\end{array}
\right.
\label{background-Sugimoto}
\end{equation}
Notice that $\Omega = e^A$ can be regarded as the scale factor of cosmological expansion,
 while $\tau$ defines implicitly the cosmic time via
\begin{equation}
 dt \ = \ e^{ - \frac{3}{4} \phi} \ \sqrt{\alpha_E} \ d\tau \ .
\end{equation}

This result of~\cite{Dudas:2000ff} is a special instance of a class of solutions
 for Einstein gravity minimally coupled to a scalar field
 in the presence of an exponential potential proportional to $\exp(\gamma\phi)$,
 which were studied in detail in~\cite{russo}.
The exponent selected by String Theory for ``brane supersymmetry breaking'',
 $\gamma = 3/2$ in the Einstein frame,
 has the key property of separating two regions of solutions with widely different behavior.
As pointed out in~\cite{climbing},
 this value marks the onset of the ``climbing phenomenon'',
 according to which the scalar field has no other option,
 when emerging from the initial singularity,
 than climbing up the potential before reaching a turning point and starting its descent.
This is important,
 since this dynamics sets naturally an upper bound on the string coupling
 (although there are no indications of a similar bound on $\alpha'$ corrections~\cite{cd}),
 and suggests that the resulting descent could help one model the onset of inflation.
A climbing phase could provide the impulse to start inflation~\cite{inflation},
 and in general an early fast--roll would introduce a low--frequency cut
 in the primordial power spectrum of scalar perturbations.
This option was explored over the years in different contexts~\cite{CMBcut},
 but it arguably obtains, in ``brane supersymmetry breaking'',
 an enticing input from String Theory.
There are also some signs, away from the Galactic plane,
 of an encouraging comparison with the lack of power apparently present
 in the low--$\ell$ CMB~\cite{comparison}.

There is an important link with the preceding case,
 since close to the turning point for the climbing scalar field,
 where $\dot{\phi}=0$ and $\ddot{A}=0$,
 the Universe in this case bears some similarities with de Sitter spacetime
 with temporally constant $\dot{A}$ in $\Omega = e^A$.
The Christoffel symbols and the spin connection for this background are
\begin{equation}
 \Gamma^0{}_{ij} = \sqrt{\alpha_E} \, \dot{A} e^{2A-2B} \delta_{ij},
\quad
 \Gamma^0{}_{00} = \sqrt{\alpha_E} \, \dot{B},
\quad
 \Gamma^i{}_{0j} = \Gamma^i{}_{j0} = \sqrt{\alpha_E} \, \dot{A} \delta^i_j,
\quad
 \mbox{others = 0},
\end{equation}
and
\begin{equation}
 \omega_\mu{}^{0{\bar m}}
  = - \omega_\mu{}^{{\bar m}0} = e^{-B} \sqrt{\alpha_E} \, \dot{A} e_\mu{}^{{\bar m}},
\qquad
 \mbox{others = 0},
\end{equation}
 respectively.
The curvature tensors are described as
\begin{eqnarray}
 R^\mu{}_{\nu\rho\sigma}
 &=& \alpha_E
     \Big[
     \delta^\mu_0 ( \delta^0_\rho \eta_{\nu\sigma} - \delta^0_\sigma \eta_{\nu\rho} )
     ( \ddot{A} - \dot{A} \dot{B} ) e^{2A - 2B}
\nonumber\\
 && + \,\, \delta^0_\nu ( \delta^0_\rho \delta^\mu_\sigma - \delta^0_\sigma \delta^\mu_\rho )
     ( \ddot{A} + \dot{A}^2 - \dot{A} \dot{B} - \dot{A}^2 e^{2A-2B} )
\nonumber\\
 && + \,\, ( \delta^\mu_\rho \eta_{\nu\sigma} - \delta^\mu_\sigma \eta_{\nu\rho} )
     \dot{A}^2 e^{2A-2B}
     \Big],
\end{eqnarray}
\begin{eqnarray}
 R_{\nu\sigma}
 &=& \alpha_E
     \Big[
     (\eta_{\nu\sigma} - \delta^0_\sigma \eta_{\nu0} )
     ( \ddot{A} - \dot{A}\dot{B} ) e^{2A-2B}
\nonumber\\
 && - \,\, (D-1) \delta^0_\nu \delta^0_\sigma
     ( \ddot{A} + \dot{A}^2 - \dot{A} \dot{B} - \dot{A}^2 e^{2A-2B} )
\nonumber\\
 && + \,\, (D-1) \eta_{\nu\sigma} \dot{A}^2 e^{2A-2B}
     \Big],
\end{eqnarray}
\begin{equation}
 R = e^{-2B} \alpha_E \left[ 2 (D-1) ( \ddot{A} - \dot{A}\dot{B} ) + D(D-1) \dot{A}^2 \right],
\end{equation}
 where in the application to the original model one should set $D=10$.

The field equations of the gravitino and the dilatino that follow
 from the Lagrangian of eq.(\ref{model-Sugimoto}) are
\begin{equation}
 \Gamma^{\mu\rho\nu} D_\rho \psi_\nu
  + \frac{1}{2\sqrt{2}} \Gamma^\nu \Gamma^\mu \lambda \partial_\nu \phi = 0,
\end{equation}
\begin{equation}
 \Gamma^\mu D_\mu \lambda
  + \frac{1}{2\sqrt{2}} \Gamma^\mu \Gamma^\nu \psi_\mu \partial_\nu \phi = 0
\end{equation}
 and are to be combined with the constraint of eq.(\ref{constraint-Sugimoto}).

Let us now try to find a solution with vanishing dilatino $\lambda=0$.
In this case, the field equation of the gravitino becomes
\begin{equation}
 \Gamma^{\mu\rho\nu} D_\rho \psi_\nu = 0
\end{equation}
 with the constraints $\Gamma^\mu \psi_\mu = 0$ from eq.(\ref{constraint-Sugimoto})
 and $\partial_\mu \phi \, \psi^\mu = 0$, which follow from the field equation of the dilatino.
Applying $\Gamma_\mu$ to the field equation
 gives the additional constraint $D_\mu \psi^\mu = 0$,
 while the covariant divergence of the field equation
 gives further additional constraint $R_{\mu\nu} \Gamma^\nu \psi^\mu = 0$.

It is simple to see that all of these four constraints are solved by
 $\psi_0 = 0$ and $\Gamma^i \psi_i = 0$, using the explicit forms of the Christoffel symbols,
 the spin connection and the Ricci tensor.
Therefore, the problem is reduces to solve the field equation
\begin{equation}
 \Gamma^\mu D_\mu \psi_i = 0,
\end{equation}
 where the field is subject to the constraint $\Gamma^i \psi_i = 0$.
The explicit form of the second order equation
 $\Gamma^\nu D_\nu \Gamma^\mu D_\mu \psi_i = 0$ is
\begin{equation}
 \ddot{\psi}_i + ( 7 \dot{A} + \frac{3}{4} \dot{\phi} ) \dot{\psi}_i
 + \frac{49}{4} \dot{A}^2 \psi_i - \frac{7}{32} \dot{\phi}^2 \psi_i = 0,
\end{equation}
 but rescaling the field according to
\begin{equation}
 \psi_i \longrightarrow \Omega^{w_{3/2}} \,\psi_i
\end{equation}
 it takes a very simple form:
\begin{equation}
 \ddot{\psi}_i + \frac{3}{4}\, \dot{\phi} \,\dot{\psi}_i = 0.
\end{equation}
Using the explicit expression of the background in eq.(\ref{background-Sugimoto}),
 the field equation reduces to
\begin{equation}
 \frac{d^2\psi_i}{d\tau^2}
  + \frac{3}{4} \left( \frac{2}{3\tau} - \frac{3\tau}{2} \right) \frac{d{\psi}_i}{d\tau} = 0,
\end{equation}
 whose solution can be expressed in terms of the incomplete $\Gamma$-function, according to
\begin{equation}
 \psi_i = C^{(1)}_i + C^{(2)}_i \, \Gamma\left(\, \frac{1}{4}, - \frac{9}{16} \tau^2 \right),
\end{equation}
 where $C^{(1)}_i$ and $C^{(2)}_i$ are integration constants
 that satisfy the conditions $\Gamma^i C^{(1)}_i = 0$ and $\Gamma^i C^{(2)}_i = 0$.
They key point is that these solutions are not oscillatory.
Moreover, if we require a finite $\psi_i$ at $\tau \rightarrow \infty$, the only option is
\begin{equation}
 \psi_i = C^{(1)}_i
\end{equation}
 with $\Gamma^i C^{(1)}_i = 0$,
 which points again to a massless gravitino in view of our new criterion,
 namely no ``rest mass'' or no contribution to the energy density $T_{00}$.
In this respect,
 we can conclude that the gravitino is massless in this cosmological vacuum,
 which somehow replaces flat spacetime in the Sugimoto model.

The next issue concerns the actual number of degrees of freedom
 that are carried by the gravitino in the unitary gauge.
To begin with, a Majorana vector--spinor field $\psi_\mu$
 has $D \times 2^{[D/2]}$ real degrees of freedom, or 320 in $D=10$.
Since $\psi^0=0$ and $\Gamma^i \psi_i=0$, this number is readily cut to 256 in $D=10$.
Moreover, $\psi_\mu$ is also a Weyl field,
 so that number is again reduced by a factor of two,
 or to 128 degrees of freedom in $D=10$.
As usual, the first order Dirac equation that is left in unitary gauge
\begin{equation}
 \Gamma^\rho D_\rho \psi_\mu = 0
\end{equation}
 amounts, in our background, to a condition $\gamma^0 \psi_\mu = 0$,
 and considering projection operators $P_{\pm} = (1 \pm i\gamma^0)/2$,
 one is finally left with 64 degrees of freedom.
This number results precisely from
 the 56 degrees of freedom carried by the original gravitino
 and the 8 degrees of freedom carried by the Nambu--Goldstone fermion.

To reiterate, the gravitino in the Sugimoto model
 behaves as a massless field in its cosmological background of~\cite{Dudas:2000ff},
 although it combines the degrees of freedom
 that a massless gravitino would describe in a flat spacetime
 with those of the absorbed Nambu--Goldstone fermion.
The two fields recover separate lives,
 consistently with this interpretation and known facts,
 if one turns off the tadpole potential.
Moreover,
 this number is the half of the degrees of freedom
 that a massive spin--$3/2$ field would have in a ten--dimensional Minkowski spacetime,
 as demanded by the orientifold projection that underlies the Sugimoto model.

\section{Conclusions}
\label{conclusions}

We have investigated the behavior of the gravitino
 in a class of orientifold models~\cite{orientifolds}
 with ``brane supersymmetry breaking''~\cite{Sugimoto:1999tx,bsb},
 where supersymmetry is non--linearly realized while a gravitino mass term is not allowed.
These models require non--trivial spacetime backgrounds,
 so that the notion of mass entails some subtleties
 and is different from the more familiar case of a flat spacetime.
In a de Sitter spacetime
 null--cone propagation in the symmetric coordinate system provides
 an accepted criterion of masslessness,
 and along these lines we have proposed a new criterion for fermions
 that applies in the flat slicing coordinate system.
The new criterion rests on a cosmological interpretation,
 and applies to more general background spacetimes.
These include the spatially flat geometries
 where ``brane supersymmetry breaking'' brings along the climbing mechanism~\cite{climbing}.
This could provide the initial impulse to start an inflationary phase~\cite{inflation},
 and with a short inflation this type of dynamics
 could have had some bearing on the low--$\ell$ CMB anomalies~\cite{comparison}.

We have investigated the gravitino mass in the Sugimoto model,
 with reference to the background field configuration of~\cite{climbing},
 which is more complicated than de Sitter spacetime,
 arriving at conclusions that are reasonable with a massless gravitino.
The overall lesson is consistent with massless gravitinos in non--trivial backgrounds
 that entail different numbers of degrees of freedom than in flat spacetime.
Let us also stress that
 the ``cosmological'' criterion of masslessness in de Sitter--like backgrounds
 also applies to a conformal scalar field theory
 with non--minimal coupling to the scalar curvature, consistently with our view.
Even in this case,
 the field equation for a suitably rescaled scalar field allows non--oscillatory solutions
 for which the energy density vanishes, although there is also a different option,
 which is proportional to $a^{-D}$, where $a=\exp(Hx^0)$ is the scale factor.
The latter contribution is typical of massless radiation.

There is a host of evidence that
 our Universe has never been exactly a flat Minkowski spacetime,
 and that now it is close to a de Sitter spacetime.
Fields with unusual numbers of degrees of freedom can thus be of interest,
 in principle, for Cosmology.
A recent investigation along these lines
 can be found in~\cite{Kehagias:2017cym,Bartolo:2017sbu,Baumann:2017jvh,Franciolini:2017ktv},
 for example.

\section*{Acknowledgments}

The author would like to thank S.~Ferrara for helpful discussions
 and A.~Sagnotti for helpful discussions, suggestions and a careful reading of the manuscript.
The author would also like to thank Scuola Normale Superiore for the kind hospitality
 and for partial support while this work was in progress.

\end{document}